\documentclass[epj]{svjour}
\usepackage{graphicx}
\input{epsf}
\usepackage{amssymb}

\begin{document}

\title{Spectral properties of Google matrix  of Wikipedia and other networks}

\author{
L. Ermann\inst{1,2}
\and
K.M. Frahm\inst{2}
\and
D.L. Shepelyansky\inst{2}
}
\institute{
Departamento de F\'\i sica Te\'orica, GIyA, Comisi\'on Nacional 
de Energ\'ia At\'omica, Buenos Aires, Argentina\\
\and
Laboratoire de Physique Th\'eorique du CNRS, IRSAMC, 
Universit\'e de Toulouse, UPS, 31062 Toulouse, France
}

\titlerunning{Google matrix of Wikipedia}
\authorrunning{L.Ermann, K.M.Frahm  and D.L.Shepelyansky}

\abstract{We study the properties of eigenvalues and eigenvectors of 
the Google matrix of the Wikipedia articles hyperlink network
and other real networks. With the help of the Arnoldi method
we analyze the distribution of eigenvalues in the complex plane
and show that eigenstates with significant eigenvalue
modulus are located on well defined network
communities. We also show that the correlator between
PageRank and CheiRank vectors distinguishes
different organizations of information flow
on BBC and Le Monde web sites. 
}

\PACS{
{89.75.Fb}{
Structures and organization in complex systems}
\and
{89.75.Hc}{
Networks and genealogical trees}
\and
{89.20.Hh}{
World Wide Web, Internet}
}

\date{Received: December 5, 2012}

\maketitle

\section{ Introduction}
With the appearance of the World Wide Web (WWW) \cite{www}
the modern society created  huge directed networks
where the information retrieval and ranking of 
network nodes  becomes a formidable challenge.
The mathematical grounds of ranking of nodes 
are based one the concept of Markov chains \cite{markov}
and related class of
Perron-Frobenius operators naturally appearing in
dynamical systems (see e.g. \cite{mbrin}).
A concrete implementation of these mathematical concepts
to the ranking of WWW nodes was started by Brin and Page
in 1998 \cite{brin}. It is significantly based on the PageRank
algorithm (PRA) which became  a fundamental element
of the Google search engine broadly used by
Internet users \cite{meyerbook}.

Already in 1998 Brin and Page pointed out that
{\it ``despite the importance of large-scale search engines on the web,
very little academic research has been done on them''} \cite{brin}.
Since that time the academic studies 
have been concentrated mainly on the properties
of the PageRank vector determined by the PRA
(see e.g. \cite{donato},\cite{upfal,litvak},\cite{meyerbook}).
Of course, the PageRank vector is at the basis 
of ranking of network nodes but the whole
description of a directed network is
given by the Google matrix $G$. 
Thus it is important to understand 
the properties of the whole spectrum of eigenvalues of Google matrix 
and to analyze the meaning and significance of its
eigenstates. Certain spectral properties of $G$ matrix
have been analyzed in 
\cite{capizzano},\cite{ggs1,ggs2},
\cite{linux,univuk},\cite{wtrade,twitter}.
Here, we concentrate our spectral analysis on the
Wikipedia articles network studied
in \cite{zzswiki}. The advantage of this
network is due to a clear meaning of nodes,
determined by the titles of Wikipedia
articles thus simplifying the understanding
of information flow in this network.
In addition to that we analyze the statistical
properties of eigenvalues and eigenstates of $G$
for WWW networks of Cambridge University, Python,
BBC and Le Monde crawled in March 2011.

The Google matrix elements of a directed network are defined as 
\cite{brin,meyerbook,gmatrixwiki}
\begin{equation}
   G_{ij} = \alpha  S_{ij} + (1-\alpha)/N \;\; ,
\label{eq1} 
\end{equation} 
where the matrix $S_{ij}$ is obtained from an adjacency matrix 
$A_{ij}$ by normalizing all nonzero columns to one ($\sum_j S_{ij}=1$) 
and replacing columns with only zero elements 
by $1/N$ ({\em dangling nodes}) with $N$ being the matrix size.
For the WWW an element $A_{ij}$ of the adjacency matrix 
is equal to unity
if a node $j$ points to the node $i$ and zero otherwise.
The damping parameter $\alpha$ in the WWW context 
describes the probability 
$(1-\alpha)$ to jump to any node for a random surfer. 
For WWW the Google search engine uses 
$\alpha \approx 0.85$ \cite{meyerbook}.
The matrix $G$ belongs to the class of Perron-Frobenius 
operators \cite{meyerbook},
its largest eigenvalue 
is $\lambda = 1$ and other eigenvalues have $|\lambda| \le \alpha$. 
The right eigenvector at $\lambda = 1$, which is called the PageRank, 
has real nonnegative elements $P(i)$
and gives a probability $P(i)$ to find a random surfer at site $i$. 
Due to the gap $1-\alpha\approx 0.15$ between the largest eigenvalue and the 
other eigenvalues the PRA permits an efficient and simple determination of the 
PageRank by the power iteration method. 
Note that at $\alpha=1$ the largest eigenvalue $\lambda=1$ is 
typically highly degenerate to due to many invariant subspaces which 
define many independent Perron-Frobenius operators which provide (at least)
one eigenvalue $\lambda=1$. This point and also a numerical method to 
determine the PageRank for the case $1-\alpha\ll 1$ are described in detail in 
\cite{univuk}. 

Once the PageRank (at $\alpha=0.85$) is found, 
all nodes can be sorted by decreasing probabilities $P(i)$. 
The node rank is then given by index $K(i)$ which
reflects the  relevance of the node $i$. The top 
PageRank nodes are located at small values of $K(i)=1,2,...$.

In addition to a given directed network $A_{ij}$
it is useful to analyze an inverse network
with inverted direction of links with
elements of adjacency matrix $A_{ij} \rightarrow A_{ji}$.
The Google matrix $G^*$ of the inverse network 
is then constructed via corresponding matrix $S^*$
according to the relations
(\ref{eq1}) using the same value of $\alpha$ as for the $G$ matrix.
The right eigenvector of $G^*$ at eigenvalue $\lambda=1$
is called CheiRank giving a complementary rank index $K^*(i)$ of network nodes
\cite{twitter,zzswiki},\cite{alik,2dmotor},\cite{cheirankwiki}.
It is known that the PageRank probability 
is proportional to the number of ingoing links
characterizing how popular or known a given node is while the 
CheiRank probability is proportional to the 
number of outgoing links highlighting the node communicativity
(see e.g. \cite{meyerbook,donato},
\cite{upfal,litvak},\cite{zzswiki,2dmotor}).
The statistical properties of the node distribution
on the PageRank-CheiRank plane are described 
in \cite{2dmotor} for various directed networks.

The paper is composed as following:
the spectrum of the Google matrix of various networks
is analyzed in Section 2, statistical properties of eigenstates
are discussed in Section 3, the communities related to
Wikipedia eigenstates are examined in Section 4,
the distribution of nodes in the PageRank-CheiRank plane is 
studied in Section 5, the link distribution over PageRank index
is considered in Section 6,
discussion of results is given in Section 7,
acknowledgments are given in Section 8,
Appendix Section 9 gives all parameters
of the 5 directed networks considered here
and describes in detail certain eigenvalues and eigenvectors.

\section{Google matrix spectrum}

We study the spectrum of eigenvalues of the Google matrix of
5 directed networks. For each network 
the number of nodes $N$ and the number of links $N_\ell$
are given in Table 1 (see Appendix). The spectrum
is obtained numerically using the powerful Arnoldi method
described in \cite{arnoldibook,golub},\cite{ulamfrahm}. The idea 
of the method is to construct a set of orthonormal vectors by applying 
the matrix ($G$, $S$, $G^*$, $S^*$ 
or any other matrix of which we want to determine 
the largest eigenvalues) on some suitable normalized initial vector and 
orthonormalizing the result to the initial vector. Then the matrix is applied 
to the second vector and the result is orgthonormalized to the first 
two vectors and so on. The used scalar products and normalization factors 
during the Gram-Schmidt process provide the matrix representation of the 
initial big matrix on the set of orthonormal vectors (which 
span a {\em Krylov space}) in a form of a 
Hessenberg matrix whose eigenvalues converge typically quite well versus 
the largest eigenvalues of the initial matrix even if the chosen number 
of orthonormal vectors, the Arnold dimension $n_A$, is quite modest 
(3000-5000 in this work) as compared to the initial matrix size. 

In this work we are interested in the spectrum of the matrix $S=G(\alpha=1)$ 
(or $S^*$) since the spectrum of $G(\alpha)$ (or $G^*(\alpha)$) is simply 
obtained by rescaling the complex eigenvalues with 
the factor $\alpha$ (apart from ``one'' largest eigenvalue $\lambda=1$
which does not change).

However, the highly degenerate unit eigenvalue $\lambda=1$ of $S$ 
creates convergence problems for the Arnoldi method and therefore as 
in \cite{univuk,twitter} we first find 
the invariant isolated subsets. 
These subsets are invariant with respect to applications of $S$.
We merge all subspaces with common 
members, and obtain a sequence of disjoint subspaces $V_j$ of dimension 
$d_j$ invariant by applications of $S$. The remaining part of nodes 
forms the wholly connected {\it core space}. 
Such a classification scheme can be efficiently implemented in a 
computer program and it provides a subdivision of network nodes in $N_c$ 
core space nodes and $N_s$ subspace nodes 
belonging to at least one of the invariant subspaces $V_j$ 
inducing the block triangular structure of matrix $S$:
\begin{equation}
\label{eq2}
S=\left(\begin{array}{cc}
S_{ss} & S_{sc}  \\
0 & S_{cc}\\
\end{array}\right)\;
\end{equation}
where $S_{ss}$ is itself composed of many small diagonal blocks for 
each invariant subspace and whose eigenvalues can be efficiently obtained 
by direct (``exact'') numerical diagonalization.

The total subspace size $N_S$, the number of independent subspaces $N_d$,
the maximal subspace dimension $d_{\rm max}$ and the number $N_1$
of $S$ eigenvalues with $\lambda=1$ are given in Table 2.
The spectrum and eigenstates of the core space $S_{cc}$
are determined by the Arnoldi method with Arnoldi dimension $n_A$ 
giving the eigenvalues $\lambda_i$ of $S_{cc}$ with largest modulus 
and the corresponding eigenvectors $\psi_j$
($G \psi_i = \lambda_i \psi_i$). The values of $n_A$ we used for 
the different networks are given in Table 1.
According to Table 2 we have the average number of links per node
$\zeta_\ell \approx 21.63$ (Wikipedia), $16.91$ (Cambridge 2011),
$16.67$ (Python), $22.77$ (BBC), $79.14$ (Le Monde).

\begin{figure}
\begin{center}
\includegraphics[clip=true,width=7.7cm]{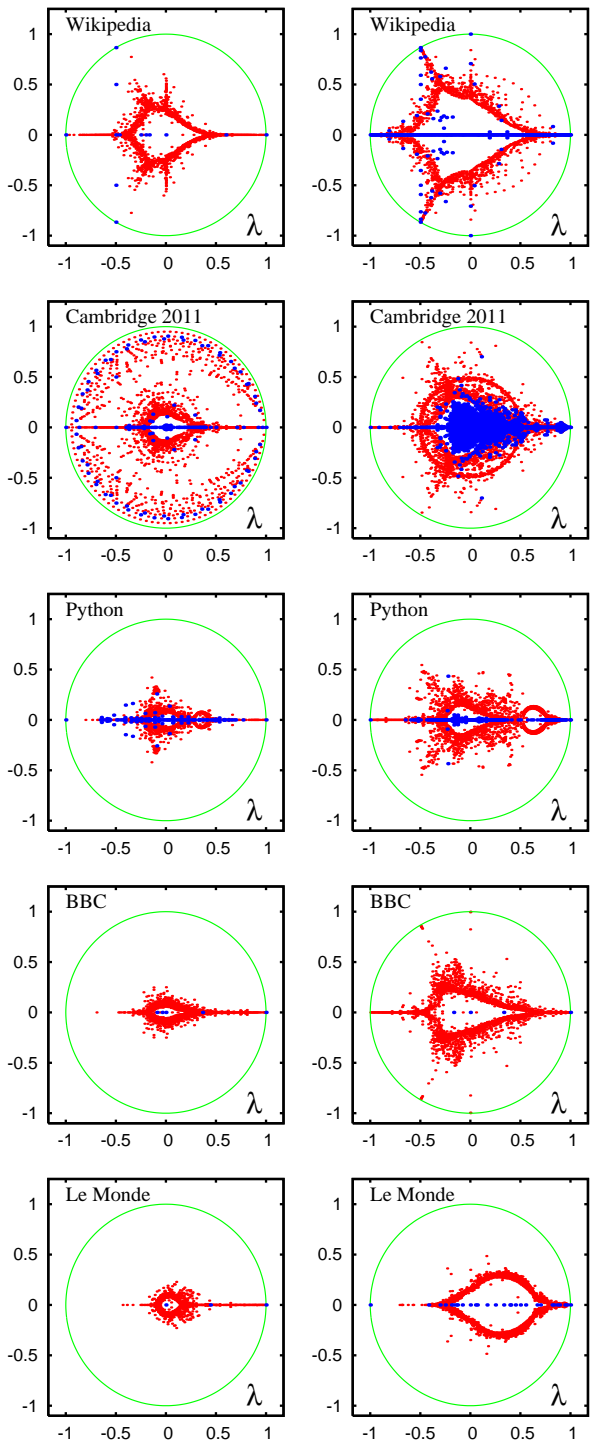}
\vglue -0.1cm
\caption{Spectrum of eigenvalues $\lambda$ the Google matrices $G$ 
(left column) and $G^*$ (right column) for Wikipedia, 
Cambridge 2011, Python, BBC and Le Monde
($\alpha=1$). Red dots are core space eigenvalues, blue dots are subspace 
eigenvalues and the full green curve shows the unit circle. 
The core space eigenvalues were calculated 
by the projected Arnoldi method with 
Arnoldi dimensions $n_A$ as given in Table 1.
}
\label{fig1}
\end{center}
\end{figure}

The distributions of subspaces eigenvalues
and largest $n_A$ eigenvalues of the core space
are  shown in Fig.~\ref{fig1}
in the complex plane $\lambda$ for all 5 networks. 
The blue points show the eigenvalues of isolated subspaces.
We note that their number is relatively small
compared to those of British University 
networks \cite{britishuniv} (up to year 2006)
analyzed in \cite{univuk}. 
We attribute this to a larger number of 
$\zeta_\ell$ links per node that reduces
an effective size of isolated parts of network. Between 2006 and 2011, 
especially for Cambridge, 
it seems that the increased use of PHP and similar web software 
tends to considerably increase the value of $\zeta_\ell$. 
Indeed, we have $\zeta_\ell \approx 10 $ for
university networks up to 2006 \cite{univuk} which used less this 
kind of PHP software.
In Fig.~\ref{fig1} the red points show $n_A$
eigenvalues of the core space  with largest $|\lambda|$.
Due to finite $n_A$ value there is an empty white space 
around $\lambda=0$. There is no significant gap 
for core eigenvalues since $\lambda_1$ is rather close to $1$ (see 
Table 3). 

In global we can say that the structure of the Wikipedia
spectrum of $S$ and $S^*$
is somewhat similar to those of Cambridge 2006 
(see Fig.2 in \cite{univuk}). For Cambridge 2011
the spectrum of $S$ is drastically changed
compared to the year 2006 but for $S^*$
certain features remain common both for 2006 and 2011
(e.g. a circle $|\lambda| \approx 0.5$, triplet-star).
For Python, BBC and Le Monde the imaginary parts ${\rm Im}(\lambda)$ of 
eigenvalues of $S$ are relatively small 
compared to the networks of Wikipedia and Cambridge.
We suppose that there are less symmetric links in the later cases.
It is interesting that for $S^*$ 
of Python, BBC and Le Monde
the imaginary parts  ${\rm Im}(\lambda)$
are significantly larger than for $S$.

The origin of nontrivial structures
of the spectrum of $G$ and $G^*$
for directed networks discussed here and in
\cite{ggs2,linux,univuk,twitter}
still require detailed analysis.
We note that well visible triplet and cross structures 
(see e.g. Wikipedia spectrum in Fig.~\ref{fig1}
and Fig.2 of \cite{univuk})
naturally appear in the 
spectra of random unistochastic matrices of size $N=3$ and $4$ which 
have been analyzed analytically and numerically in \cite{karol2003}.
In view of this similarity we suppose that
networks with such structures have some triplet or quartet
subgroup of nodes weakly coupled to the rest of the network.
However, a detailed understanding of the spectrum
requires a deeper analysis. In the next Section we turn to a study 
of eigenstate properties.

\section{Statistical properties of  eigenstates}

\begin{figure}
\begin{center}
\includegraphics[clip=true,width=7.7cm]{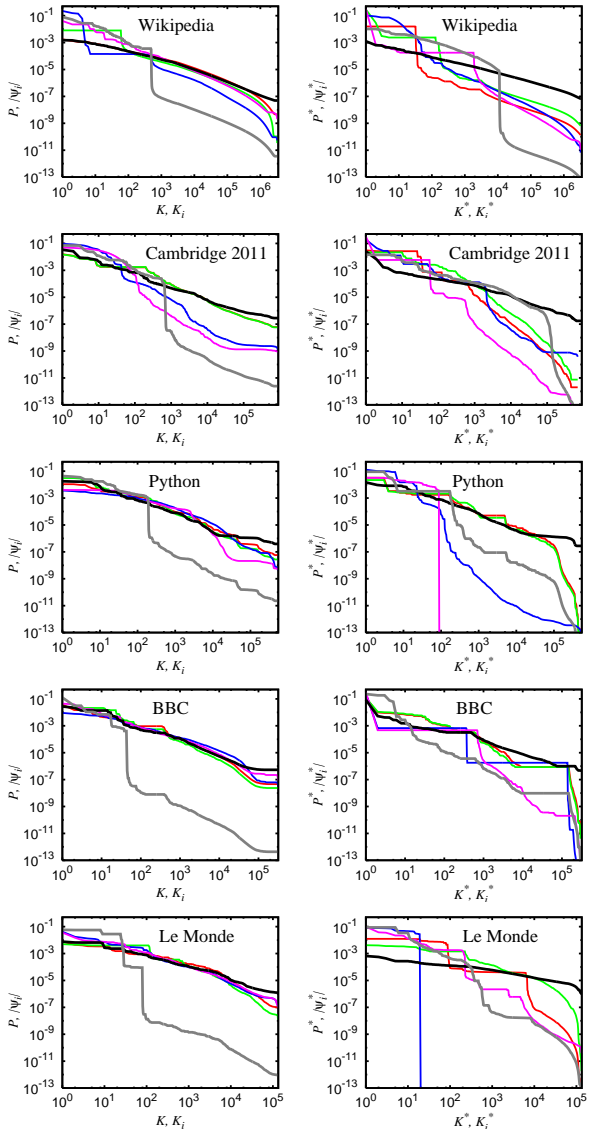}
\vglue -0.1cm
\caption{PageRank $P$ (left column) and CheiRank $P^*$ (right column) 
vectors are shown as a function of 
the corresponding rank indexes $K$ or $K^*$ 
for the Google  matrices of Wikipedia, Cambridge 2011, 
Python, BBC and Le Monde at the damping parameter 
$\alpha=0.85$ (thick black curve) and $\alpha=1-10^{-8}$ 
(thick gray curve). The thin color curves show for each panel 
the modulus of four core space eigenvectors $|\psi_i|$ of 
$S$ (left column) and $|\psi_i^*|$ of $S^*$ (right column) 
versus their ranking indexes $K_i$ or $K_i^*$. 
Red and green curves are the eigenvectors corresponding to 
the two largest core space eigenvalues (in modulus) which are real and 
close to 1; blue and pink curves 
are the eigenvectors corresponding to two complex eigenvalues with 
large imaginary part. The chosen eigenvalues and other relevant 
quantities for each case are listed 
in Tables 1,\,2,\,3.
}
\label{fig2}
\end{center}
\end{figure}

The dependence of 
PageRank  $P$ and CheiRank  $P^*$ vectors 
on their indexes $K$ and $K^*$ at
$\alpha=0.85; 1-10^{-8}$ are shown in Fig.~\ref{fig2}.
At $\alpha=0.85$ we have an approximate
algebraic decay of probability according to the Zipf law
$P \sim 1/K^\beta, P^* \sim 1/{K^*}^\beta$
(see e.g. \cite{wtrade} and Refs. therein). 
We find the following values $\beta$ for PageRank (CheiRank):
$0.96 \pm 0.002 (0.73 \pm 0.003)$ Wikipedia;
$0.81 \pm 0.007  (0.90 \pm 0.004)$ Cambridge 2011;
$1.12 \pm 0.01 (1.17 \pm 0.006)$ Python;
$1.20 \pm 0.006 (0.96 \pm 0.004)$ BBC;
$1.08 \pm 0.009 (0.55 \pm 0.002)$ Le Monde.
Formally, the statistical errors in $\beta$ are
relatively small but in some cases
there are variations of slope in the decay of PageRank
(CheiRank) probability that gives 
a dependence of $\beta$ on a fitting range
(e.g. that's why $\beta$ here is a bit different from
its values for Wikipedia given in \cite{zzswiki}). 
We note that the value $\beta \approx 1$ for the PageRank 
remains relatively stable to all networks
corresponding to the usual exponent $\mu \approx 2.1$
of algebraic decay of the ingoing link distribution
leading to $\beta=1/(\mu-1) \approx 0.9$ (see e.g. \cite{donato,upfal},
\cite{zzswiki,wtrade,twitter}). 

For CheiRank the variations
of $\beta$ from one network to another are
more significant being in agreement with the fact
that for outgoing links the exponent $\mu \approx 2.7$
varies in a more significant manner.

For $\alpha=1-10^{-8}$ we find that the main probability
of PageRank and CheiRank eigenvectors is located on isolated
subspaces with $N_s$ nodes; after that value there is a significant
drop of probability for $K,K^* > N_s$.
This effect was already found and explained in detail in \cite{univuk}
and our new data confirms that it is indeed rather generic.

\begin{figure}
\begin{center}
\includegraphics[clip=true,width=7.7cm]{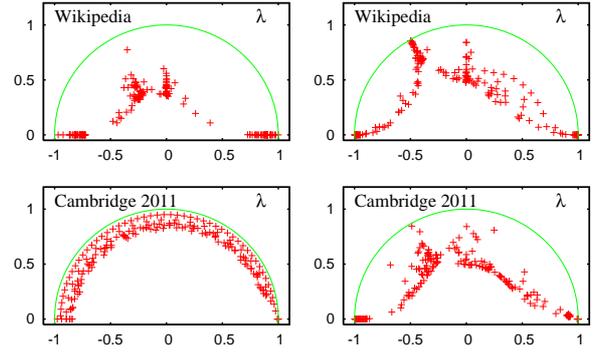}
\vglue -0.1cm
\caption{A selection of 200 complex core space eigenvalues 
closest to the unit circle 
for the matrices $S$ 
(left column) and $S^*$ (right column) of Wikipedia 
and Cambridge 2011 networks.
The characteristics of corresponding eigenvectors
are shown in Figs.~\ref{fig4},\ref{fig5}.
}
\label{fig3}
\end{center}
\end{figure}

The modulus of four eigenfuctions 
$|\psi_i(j)|$
from the core space are shown
in Fig.~\ref{fig2} by color curves
as a function of their own 
index $K_i$ which order
$|\psi_i(j)|$ in a monotonic decreasing order.
For Python, BBC and Le Monde the decay of $|\psi_i(j)|$
with $K_i$ is similar to the decay of PageRank probability with $K$.
For Wikipedia and Cambridge 2011 we see that
eigenvectors $|\psi_i(j)|$ are more localized.
The eigenstates of $S^*$ have a significantly more
irregular decay compared to the eigenstates of $S$.

\begin{figure}
\begin{center}
\includegraphics[clip=true,width=7.7cm]{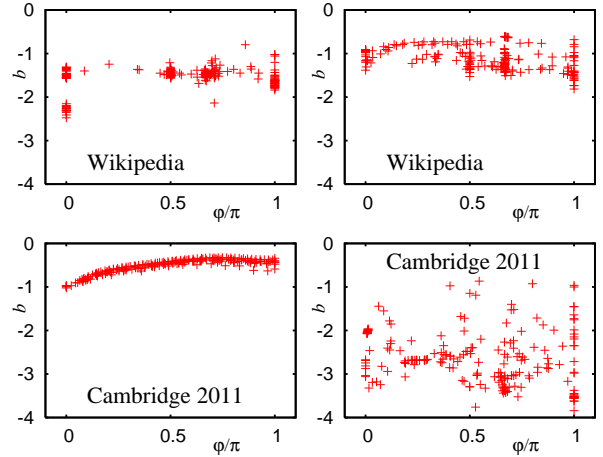}
\vglue -0.1cm
\caption{{\em Left column:} Algebraic exponent $b$ obtained from a 
power law fit 
$|\psi_i(K_i)|\sim \,K_i^b$ for $K_i\ge 10^4$ 
shown as a function of  the phase 
$\varphi=\arg(\lambda_i)$  of the complex eigenvalue $\lambda_i$ 
associated to the eigenvector $\psi_i$ of $S$. 
The shown data points correspond to the 
eigenvalue selection of Fig.~\ref{fig3} 
for  networks of Wikipedia and Cambridge 2011. 
{\em Right column:} The same as in the left column for the 
eigenvectors of $S^*$.
}
\label{fig4}
\end{center}
\end{figure} 

\begin{figure}
\begin{center}
\includegraphics[clip=true,width=7.7cm]{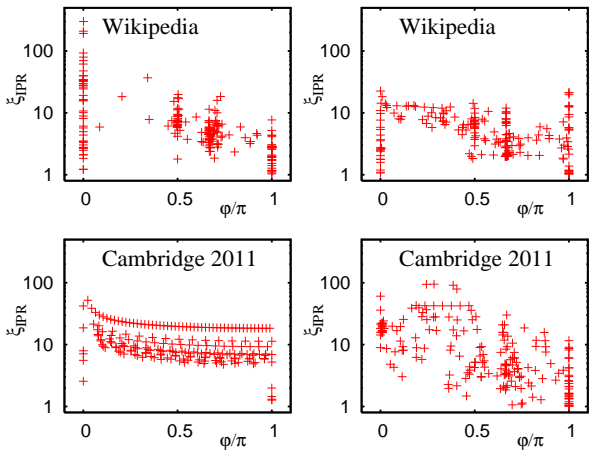}
\vglue -0.1cm
\caption{{\em Left column:} Inverse participation ratio 
$\xi_{\rm IPR}=(\sum_j |\psi_i(j)|^2)^2/\sum_j |\psi_i(j)|^4$ 
shown as a function of  the phase 
$\varphi=\arg(\lambda_i)$ 
of the complex eigenvalue $\lambda_i$ 
associated to the eigenvector $\psi_i$ of $S$. 
The  data points correspond to the 
eigenvalue selection of Fig.~\ref{fig3} for  networks of 
Wikipedia and Cambridge 2011. 
{\em Right column:} The same as in the left column for the 
eigenvectors of $S^*$.
}
\label{fig5}
\end{center}
\end{figure}

To analyze the properties of core eigenstates of \\
Wikipedia and Cambridge 2011 in a better way we select
200 core space eigenvalues of $S$ and $S^*$ being most close to the unitary
circle $|\lambda|=1$. These eigenvalues are shown in Fig.~\ref{fig3}.
For these eigenvalues we compute the corresponding eigenvectors
$\psi_i(j)$ and by fitting a power law dependence 
$|\psi_i(K_i)|\sim \,K_i^b$ at $K_i \ge 10^4$
we determine the dependence of the exponent $b$
on the phase of the eigenvalue $\varphi=\arg(\lambda_i)$.
For Wikipedia we have values of $|b|$ distributed mainly in the range
$(1,2)$ for $S$ and in the range $(0.5,1.5)$ for $S^*$.
For Cambridge 2011 we have a more compact range  $(0.5,1)$
for $S$ while for $S^*$ there is a very broad variation of 
$|b|$ values in the range $(1,4)$.

The above approximate power law 
description of the eigenstate decay characterizes their behavior
at large $K$ values. The behavior at
low $K$ values can $\;$ be $\;$ characterized $\;$ by the inverse $\;$
participation ratio $\;$ (IPR)  $\;$
$\xi_{\rm IPR}=(\sum_j |\psi_i(j)|^2)^2/\sum_j |\psi_i(j)|^4$
which gives an approximate number of nodes on which the main probability of
an eigenstate $\psi_i(j)$ is located. We note that such a characteristic
is broadly used in disordered mesoscopic systems
allowing to detect the Anderson transition from
localized phase with finite $\xi$ to delocalized phase with
$\xi$ value comparable with the system size \cite{mirlin}.
The IPR data are presented in Fig.~\ref{fig5}
for eigenvalues selection of Fig.~\ref{fig3}.
We find that $\xi_{\rm IPR}$ values are by a factor 
$10^4$ to $10^5$ smaller then the network size $N$.
This means that these eigenstates are well localized
on a restricted number of nodes.
We try to analyze what are these nodes in next Section for the 
example of Wikipedia where the meaning of a 
node is clearly defined by the title of the corresponding Wikipedia article.
 
\section{Communities of Wikipedia eigenstates}

\begin{figure}
\begin{center}
\includegraphics[clip=true,width=8.2cm]{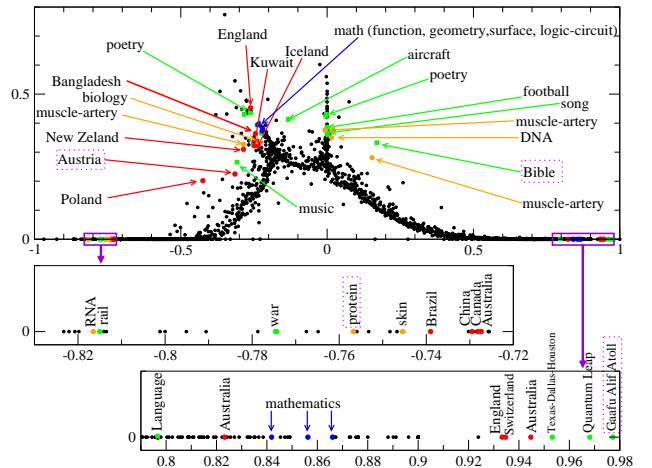}
\vglue -0.1cm
\caption{Complex eigenvalue spectrum of the matrices $S$ for Wikipedia.
Highlighted eigenvalues represent different communities 
of Wikipedia and are labeled by the most 
repeated and important words following word counting of first 1000 nodes.
Color are used in the following way: red for countries, orange 
for biology, blue for mathematics and green for others.
Top panel shows complex plane for positive imaginary part
of eigenvalues, 
while middle and bottom panels focus in 
the negative and positive real parts. Top 20 nodes with largest values of 
eigenstates $|\psi_i|$ and their eigenvalues $\lambda_i$
are given in Tables 4,\,5,\,6,\,7 
(4 names marked by dotted boxes in figure panels).
}
\label{fig6}
\end{center}
\end{figure}

To understand the meaning of other eigenstates in the core space
we order selected eigenstates by their decreasing
value $|\psi_i(j)|$ and apply a frequency analysis 
on the first $1000$ articles with $K_i \leq 1000$.
The mostly frequent word of a given eigenvector  is
used to label the eigenvector name. These labels with 
corresponding eigenvalues are shown in Fig.~\ref{fig6}
in $\lambda-$plane. We identify four main categories
for the selected eigenvectors shown by different colors in Fig.~\ref{fig6}:
countries (red), biology and medicine (orange),
mathematics (blue) and others (green). The category of others
contains rather diverse articles about poetry, Bible, football, music,
American TV series (e.g. Quantum Leap), small geographical
places (e.g. Gaafru Alif Atoll). Clearly these eigenstates
select certain specific communities which are relatively weakly
coupled with the main bulk part of  Wikipedia that
generates relatively large modulus of $|\lambda_i|$.
The top 20 articles of eigenstate PageRank index $K_i$ are 
listed in Tables 4,\,5,\,6,\,7. 

The eigenvector of Table 4
has  a positive real $\lambda$ and 
is linked to  the main article {\it Gaafu Alif Atoll}
which in its turn is linked mainly to atolls in this region.
Clearly this case represents well localized community of 
articles mainly linked between themselves that gives
slow relaxation rate of this eigenmode with $\lambda=0.9772$ 
being rather close to unity.
 
In Table 5 we have an eigenvector
with real negative eigenvalue $\lambda=-0.8165$
with the top node {\it Photoactivatable fluorescent protein}.
This node is linked to {\it Kaede (protein)}
and {\it Eos (protein)} with the later being isolated
from coral. Its picture
is listed in {\it Portal:Berkshire/Selected picture}
which has pictures of {\it St Paul’s Cathedral}
and {\it Legoland Windsor}
that generates appearance of these,
on a first glance unrelated articles,
to be present in this eigenvector.
Thus, this eigenvector also highlights a
specific community which is somewhat stronger
coupled to the global Wikipedia core,
due to a link to selected pictures,
with a smaller modulus of $\lambda$
compared to the case of Table 4.

The eigenvector of Table 6 has a complex eigenvalue
with $|\lambda|=0.3733$ and the top article
{\it Portal:Bible}. The top three
articles of this eigenvector
have very close values of $|\psi_i(j)|$
that seems to be the reason why
we have $\varphi=\arg(\lambda_i) = \pi \cdot 0.3496$
being very close to $\pi/3$.
The Bible is strongly linked to various aspects of 
human society that leads to a relatively small
modulus value of this well defined community. 

In Table 7 we have an eigenvector which starts from the
article {\it Lower Austria} with
the eigenvalue modulus $|\lambda|= 0.3869$.
This article is linked to such articles as {\it Austria}
and {\it Upper Austria} with historical links to {\it Styria}.
It also links to its city capital 
{\it Krems an der Donau}.
The articles  {\it World War II} and {\it Jew} 
appear due to a sentence 
``Before World War II, Lower Austria 
had the largest number of Jews in Austria.''
Due to links with very popular nodes
the eigenvector of this community
has a relative small modulus of $\lambda$.

The above analysis shows that the eigenvectors
of the Google matrix of Wikipedia 
clearly identify certain communities
which are relatively weakly connected with the
Wikipedia core when the modulus of corresponding
eigenvalue is close to unity. For moderate
values of $|\lambda|$ we still have 
well defined communities which are however
have stronger links with some popular articles
(e.g. countries) that leads to a more rapid decay
of such eigenmodes. 

The above results show that the 
analysis of eigenvectors highlights
interesting features of communities and
network structure. However, a priori it is not evident
what is a correspondence between
the numerically obtained eigenvectors  
and the specific community features 
in which someone has a specific interest.
It is possible that for a well defined community 
it can be useful to construct a personalized Google matrix
(see e.g. {\cite{meyerbook}) and to perform analysis of its 
eigenstates. 

\section{CheiRank versus PageRank plane}

As it is discussed in \cite{alik},\cite{twitter,zzswiki},\cite{2dmotor}
it is useful to look on the distribution of network nodes
on PageRank-CheiRank plane $(K,K^*)$. For Wikipedia
a large scale distribution is analyzed in \cite{zswiki,2dmotor} and 
the networks of British Universities, Linux Kernel and Twitter are
considered in \cite{2dmotor} and \cite{twitter}. 

\begin{figure}
\begin{center}
\includegraphics[clip=true,width=7.7cm]{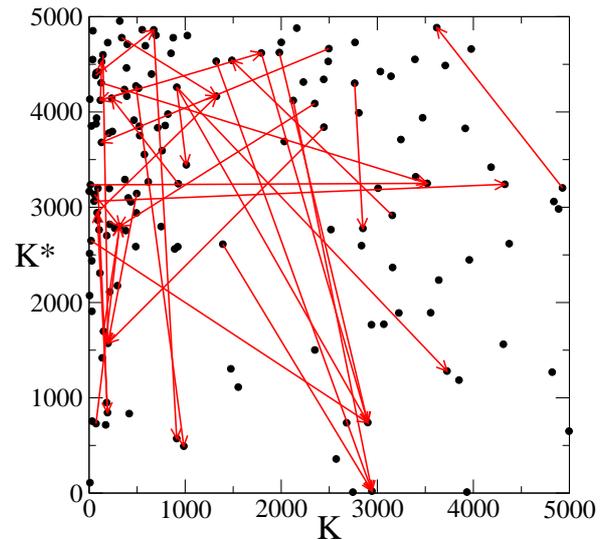}
\vglue -0.1cm
\caption{Top 5000 values in PageRank-CheiRank plane $(K,K^*)$ 
of  Wikipedia. 
All nodes and all links in this region
are shown by black circles and red  arrows respectively.
}
\label{fig7}
\end{center}
\end{figure}

In Fig.~\ref{fig7} we show for Wikipedia the distribution of
nodes in $(K,K^*)$ plane for a relatively small range
of top $5000$ values of $K,K^*$. All directed links 
in this region are also shown. In fact the number 
of such links and number of nodes in this region
are relatively small. Indeed, a large scale density of nodes
(see Fig.3 in \cite{zzswiki}) shows that the density of nodes
is not very high at the top corner of PageRank-CheiRank plane.
This happens due to the fact that top nodes of PageRank,
whose components are proportional to the number of ingoing links,
are usually not those of CheiRank, whose components are proportional to 
the number to outgoing links.

\begin{figure}
\begin{center}
\includegraphics[clip=true,width=8.2cm]{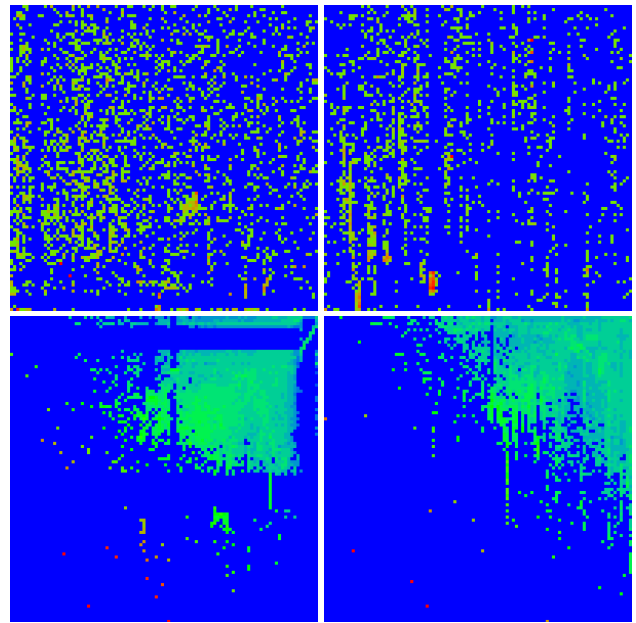}
\vglue -0.1cm
\caption{Density of nodes  $W(K,K^*)$ on PageRank-CheiRank plane $(K,K^*)$
for the networks of BBC (left panels) and Le Monde (right panels).
Top panels show density in the range $1 \leq K,K^* \leq 10^4$
with averaging over cells of size $100\times100$;
bottom panels show density averaged over $100\times100$ 
logarithmically equidistant grids for $0 \leq \ln K, \ln K^* \leq \ln N$,
the density is averaged over all nodes inside each cell of the grid,
the normalization condition is $\sum_{K,K^*}W(K,K^*)=1$.
Color varies from blue at zero value to red at maximal density value.
At each panel the $x$-axis corresponds to $K$ (or $\ln K$ for the 
bottom panels) 
and the $y$-axis to $K^*$ (or $\ln K^*$ for the bottom panels). 
}
\label{fig8}
\end{center}
\end{figure}

The correlation between PageRank and CheiRank vectors 
can be characterized by their correlator \cite{alik,2dmotor}:
\begin{equation}
  \kappa =N \sum^N_{i=1} P(K(i)) P^*(K^*(i)) - 1 \;\; .
\label{eq3} 
\end{equation}
For our networks we find its values to be $\kappa=4.08$ \\
(Wikipedia), $ 41.5$ (Cambridge 2011),
$12.9$ (Python), $ 140.2$ (BBC), $0.85$ (Le Monde).
Except for the case of Le Monde, these values are relatively 
high showing that there is a significant correlation between
PageRank and CheiRank probabilities on 
corresponding networks. We remind that for Linux Kernel networks
the values of $\kappa$ are close to zero
corresponding to absence of correlations there \cite{alik,2dmotor}.

The strong difference between $\kappa$ values for BBC and Le Monde
shows that the structure of these two web sites is very different.
To analyze this difference in a better way we show the density of nodes
for these two networks on small and large scales in Fig.~\ref{fig8}.
For small scale, shown by top panels, it is clear that
the density of nodes is significantly larger for BBC network.
However, this difference becomes even more drastic on 
the large  logarithmic scale of the whole network
shown in bottom panels.  Indeed, on a logarithmic scale we see that
BBC network has a square like distribution region
with a certain probability maximum around the diagonal $K \approx K^*$
while Le Monde network has a triangular type distribution
which is typical for networks without
correlations between PageRank and CheiRank vectors,
like it is the case for the Linux Kernel networks
(see Fig.4 in \cite{2dmotor}). Indeed, a random  procedure
of node generation on $(K,K^*)$ plane gives such a triangular
distribution without correlations between PageRank and CheiRank nodes
(see procedure description and right panel of Fig.4 in \cite{zzswiki}).
This analysis shows that BBC and Le Monde agencies handle 
information flows on their web sites in a drastically different manner.
Thus for the BBC web site the most popular articles are at the same time 
also the most communicative ones while in contrast to that 
for the Le Monde web site the most popular and most communicative articles 
are very different.

\section{Links distribution over PageRank nodes}

\begin{figure}
\begin{center}
\includegraphics[clip=true,width=7.9cm]{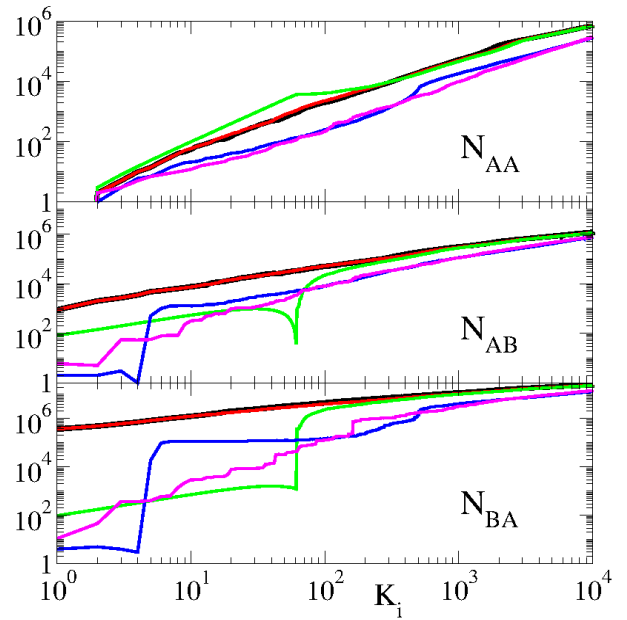}
\vglue -0.1cm
\caption{Number of links between or inside sets $A$ and $B$ 
defined by the index $K_i$ ordered by 
decreasing absolute value of Wikipedia eigenstates.
The number of links starting and pointing to nodes 
inside the set $A$ ($N_{AA}$) 
is shown in top panel as a function of $K_i$.
The cases of links from set $A$ to set $B$ ($N_{AB}$) and 
from $B$ to $A$ ($N_{BA}$) are shown in middle and 
bottom panel respectively.
Note that the total number of links is conserved and the quantity $N_{BB}$ 
can be obtained as $N_{BB}=N_\ell-N_{AA}-N_{AB}-N_{BA}$.
The case of PageRank vector with damping parameter $\alpha=0.85$ is shown 
by a black curve versus $K$ index.
The color curves show the cases of four core space 
eigenvectors $\vert\psi_i\vert$ 
of $S$ versus their ranking indexes $K_i$.
Red and green curves are the eigenvectors corresponding to 
the two largest core space
eigenvalues (in modulus) being
$\lambda_1 = 0.99998702$ and $\lambda_2 = 0.97723699$ respectively; 
blue and pink curves are the
eigenvectors corresponding to two complex eigenvalues 
with large imaginary part being 
$\lambda_{52}=-0.35003316+i0.77373677$ and 
$\lambda_{864}=-0.34293502+i0.43144930$ respectively.
}
\label{fig9}
\end{center}
\end{figure}

To understand the properties of directional flow on a network
it is also useful to analyze the distribution of links over
PageRank nodes. We illustrate this approach for the Wikipedia network.
Suppose that all nodes are ordered 
in a decreasing order of modulus of a given eigenvector.
For the PageRank vector all nodes are numbered by the PageRank index $K$,
while for a given eigenstate $\psi_i(j)$ all nodes are
numbered by a local corresponding index $K_i$.
We now divide all nodes on two parts $A$ and $B$
with $1,...,K_i$ nodes for $A$ and
$K_i+1,...,N$ nodes for $B$.
Then we determine the  number of links $N_{AA}$
starting and ending in part $A$,
the number of links $N_{AB}$ pointing from 
part $A$ to part $B$ and 
the  number of links $N_{BA}$ pointing from 
part $B$ to part $A$. The number of links inside part $B$
is then $N_{BB}=N_\ell-N_{AA}-N_{AB}-N_{BA}$.
For the PageRank vector the dependence of $N_{AA}$ on $K$
was analyzed for different networks in \cite{twitter}.
Here we generalize this concept to
consider links between two parts $A,B$
for various eigenvectors of the Google matrix.

According to the data of Fig.~\ref{fig9}
we find that for all eigenvectors
$N_{AA} \propto K_i^{1.5}$ grows
approximately in an algebraic way with the exponent 
being close to $1.5$ being similar to the PageRank case
considered in \cite{twitter}.
However, the dependence of $N_{AB}$ and $N_{BA}$
on $K_i$ is rather different for different eigenstates.
For the PageRank and the $\lambda_1$ eigenvector
we find practically the same behavior 
linked to the fact that at $\alpha=0.85$
the PageRank vector is rather close to
the first core space eigenvector (see discussion at \cite{univuk}).
Here the interesting point is that at small values of $K_i$
we have $N_{BA}$ being larger than $N_{AB}$ almost by a factor $100$.
This is due to the fact that low rank nodes at large $K_i$ point 
preferentially to high rank nodes at low $K_i$.
For other three eigenvectors with $\lambda_2$,
$\lambda_{52}$, $\lambda_{864}$ we find well pronounced step-like
behavior of  $N_{AB}, N_{BA}$ on $K_i$. 
We argue that the step size  in $K_i$
is given by the size of a community which has
preferential links mainly inside the community.
Indeed, for the eigenvector of $\lambda_2$ (see Table 3) we see
that the community size is approximately 
$N_{\rm cs} \approx 1/|\psi_1| \approx 100$ that corresponds to the 
step size in $K_i \approx 70$ for this case.  

These results show that the analysis of the link distribution
over the PageRank index provides interesting and useful
information about characteristics and properties of
directed networks.

\section{Discussion}

In this work we performed a spectral analysis of 
eigenvalues and eigenstates of the 
Google matrix of Wikipedia and other networks.
Our study shows that the spectrum of the core space component
has eigenvalues in a close vicinity of $\lambda=1$
and that there are isolated subspaces which
give a degeneracy of the eigenvalue $\lambda=1$.
The eigenvalues and eigenstates with relatively large 
values of $|\lambda|$ can be efficiently determined
by the powerful Arnoldi method.
These eigenstates are mainly located on  well defined 
network communities. We also find that the spectrum
changes drastically from one network to another
even if the distribution of links and decay of PageRank
is rather similar for the networks considered.
This means that the properties of directed networks
strongly depend on the internal network structure.
We show that the correlation between PageRank and CheiRank vectors
highlights specific properties of information
flow on directed network.
For example, this correlation 
demonstrates a  drastic difference 
between web sites of BBC and Le Monde.
The distribution of links between PageRank nodes
also provides an interesting information
about the network structure.
On the basis of our studies 
we argue that the developed spectral analysis
of Google matrix brings a deeper understanding of
information flow on real directed networks.

\section{Acknowledgments}
We thank A.D.Chepelianskii for 
making to us available network data collected by him
for networks of Cambridge University, Python, BBC, Le Monde in March 2011.
Our research presented here is supported in part by the EC FET Open project 
``New tools and algorithms for directed network analysis''
(NADINE $No$ 288956). This work was granted access to the HPC resources of 
CALMIP (Toulouse) under the allocation 2012-P0110.

\renewcommand{\theequation}{A-\arabic{equation}}
  \setcounter{equation}{0}  
\renewcommand{\thefigure}{A-\arabic{figure}}
  \setcounter{figure}{0}  
\section{Appendix}

We give here tables with all network parameters used in the paper.
The notations used in the tables are:
$N$ is network size, $N_\ell$ is the number of links,
$n_A$ is the Arnoldi dimension used for the Arnoldi method for the 
core space eigenvalues, $N_d$ is the number of invariant subspaces,
$d_{\rm max}$ gives a maximal subspace dimension,
$N_{\rm circ.}$ notes number of eigenvalues on the unit circle with 
$|\lambda_i|=1$, $N_1$ notes 
number of unit eigenvalues with $\lambda_i=1$.
We remark that $N_s\ge N_{\rm circ.}\ge N_1\ge N_d$ and $N_s\ge d_{\rm max}$ 
and the average subspace dimension is given by: $\langle d\rangle = N_s/N_d$.
We note that the values of $N$, $N_\ell$ for network of Cambridge 2011
are slightly different from those given in \cite{2dmotor}
due to a slightly different procedure of cleaning of
row data collection (e.g. count of pdf and other type nodes).
Eigenvalues for eigenvectors are shown in Fig. 1 with 
the colors red, green, blue or pink 
corresponding to colors of Table 3.
The index $m$ of $\lambda_m$ in Tables 3,4,5,6,7 counts
the order number of core eigenvalues 
in a decreasing order of $|\lambda_m|$.


\medskip
\begin{table}\label{table1}
\centering
\begin{tabular}{|l|c|c|c|}
\hline
& $N$ & $N_\ell$ & $n_A$ \\
\hline
\hline
\ Wikipedia\ & \ 3282257\ & \ 71012307\ & \ 3000\ \\
\hline
\ Cam. 2011\ & \ 893176\ & \ 15106706\ & \ 4000\ \\
\hline
\ Python\ & \ 541545\ & \ 9031262\ & \ 5000\ \\
\hline
\ BBC\ & \ 319637\ & \ 7278258\ & \ 4000\ \\
\hline
\ Le Monde\ & \ 134196\ & \ 10621445\ & \ 5000\ \\
\hline
\end{tabular}
\caption{Parameters of all networks considered in the paper.}
\end{table}

\medskip
\begin{table}\label{table2}
\centering
\begin{tabular}{|l|c|c|c|c|c|}
\hline
& $N_s$ & $N_d$ & $d_{\rm max}$ & $N_{\rm circ.}$ & $N_1$\\
\hline
\hline
\ Wikipedia\ & \ 515\ & \ 255\ & \ 11\ &\ 381\ & \ 255\ \\
\hline
\ Wikipedia$^*$\ & \ 21198\ & \ 5355\ & \ 717\ &\ 8968\ & \ 5365\ \\
\hline
\ Cam. 2011\ & \ 808\ & \ 329\ & \ 74\ &\ 343\ & \ 332\ \\
\hline
\ Cam. 2011$^*$\ & \ 186062\ & \ 2039\ & \ 5144\ &\ 2044\ & \ 2041\ \\
\hline
\ Python\ & \ 198\ & \ 23\ & \ 72\ &\ 26\ & \ 23\ \\
\hline
\ Python$^*$\ & \ 1589\ & \ 25\ & \ 951\ &\ 35\ & \ 31\ \\
\hline
\ BBC\ & \ 50\ & \ 19\ & \ 28\ &\ 19\ & \ 19\ \\
\hline
\ BBC$^*$\ & \ 39\ & \ 28\ & \ 6\ &\ 28\ & \ 28\ \\
\hline
\ Le Monde\ & \ 83\ & \ 64\ & \ 18\ &\ 64\ & \ 64\ \\
\hline
\ Le Monde$^*$\ & \ 789\ & \ 354\ & \ 15\ &\ 373\ & \ 361\ \\
\hline
\end{tabular}
\caption{$G$ and $G^*$ eigespectrum parameters for all networks.}
\end{table}

\medskip
\begin{table}\label{table3}
\centering
\begin{tabular}{|c|l|l|}
\hline
& color & \ \ \ eigenvalue \\
\hline
\hline
Wikipedia& red & $\lambda_{1}=0.999987$ \\
\hline
 & green & $\lambda_{2}=0.977237$ \\
\hline
 & blue & $\lambda_{52}=-0.35003+i\,0.77374$ \\
\hline
 & pink & $\lambda_{864}=-0.34293+i\,0.43145$ \\
\hline
Wikipedia$^*$& red & $\lambda_{1}=0.999982$ \\
\hline
& green & $\lambda_{2}=0.999902$ \\
\hline
& blue & $\lambda_{662}=0.0000000+i\,0.84090$ \\
\hline
& pink & $\lambda_{38}=-0.49626+i\,0.85653$ \\
\hline
Cam. 2011& red & $\lambda_{1}=0.999749$ \\
\hline
& green & $\lambda_{2}=0.999270$ \\
\hline
& blue & $\lambda_{350}=0.41779+i\,0.77856$ \\
\hline
& pink & $\lambda_{144}=-0.52909+i\,0.78693$ \\
\hline
Cam. 2011$^*$& red & $\lambda_{1}=0.999998$ \\
\hline
& green & $\lambda_{2}=0.999994$ \\
\hline
& blue & $\lambda_{765}=0.24846+i\,0.80915$ \\
\hline
& pink & $\lambda_{249}=-0.48736+i\,0.84568$ \\
\hline
Python& red & $\lambda_{1}=0.999975$ \\
\hline
& green & $\lambda_{2}=0.998864$ \\
\hline
& blue & $\lambda_{3315}=0.14484+i\,0.19215$ \\
\hline
& pink & $\lambda_{1337}=-0.14427+i\,0.42051$ \\
\hline
Python$^*$& red & $\lambda_{1}=0.999995$ \\
\hline
& green & $\lambda_{2}=0.999991$ \\
\hline
& blue & $\lambda_{2559}=0.37694+i\,0.45231$ \\
\hline
& pink & $\lambda_{3076}=0.12214+i\,0.47416$ \\
\hline
BBC& red & $\lambda_{1}=0.99883$ \\
\hline
& green & $\lambda_{2}=0.99251$ \\
\hline
& blue & $\lambda_{1276}=-0.12414+i\,0.24795$ \\
\hline
& pink & $\lambda_{1148}=-0.22459+i\,0.20024$ \\
\hline
BBC$^*$& red & $\lambda_{1}=0.999999$ \\
\hline
& green & $\lambda_{2}=0.999994$ \\
\hline
& blue & $\lambda_{16}=-0.00067+i\,0.99930$ \\
\hline
& pink & $\lambda_{90}=-0.49635+i\,0.85848$ \\
\hline
Le Monde& red & $\lambda_{1}=0.998837$ \\
\hline
& green & $\lambda_{2}=0.983123$ \\
\hline
& blue & $\lambda_{926}=0.10295+i\,0.22890$ \\
\hline
& pink & $\lambda_{1118}=0.08023+i\,0.20595$ \\
\hline
Le Monde$^*$& red & $\lambda_{1}=0.999999$ \\
\hline
& green & $\lambda_{2}=0.999959$ \\
\hline
& blue & $\lambda_{2093}=0.15987+i\,0.48502$ \\
\hline
& pink & $\lambda_{2474}=0.17637+i\,0.40917$ \\
\hline
\end{tabular}
\caption{Eigenvalues of eigenvectors shown in Figs.~\ref{fig1},\ref{fig2} by corresponding colors. Index $m$ of $\lambda_m$ numbers eigenvalues
in the decreasing order of $|\lambda|$ in the core space.}
\end{table}

\medskip
\begin{table}\label{table4}
\centering  
\begin{tabular}{|l|c|c|}
\hline
	&	$\lambda_2=0.9772$	(``Gaafu Alif Atol'') &	$\vert\psi_i\vert$	\\
\hline
\hline
1	&	Gaafu Alif Atoll 	&	0.00816	\\
2	&	Kureddhoo (Gaafu Alif Atoll)	&	0.00812	\\
3	&	Hithaadhoo (Gaafu Alif Atoll)	&	0.00808	\\
4	&	Dhigurah (Gaafu Alif Atoll)	&	0.00806	\\
5	&	Maarandhoo (Gaafu Alif Atoll)	&	0.00806	\\
6	&	Hulhimendhoo (Gaafu Alif Atoll)	&	0.00805	\\
7	&	Araigaiththaa	&	0.00798	\\
8	&	Baavandhoo	&	0.00798	\\
9	&	Baberaahuttaa	&	0.00798	\\
10	&	Bakeiththaa	&	0.00798	\\
11	&	Beyruhuttaa	&	0.00798	\\
12	&	Beyrumaddoo	&	0.00798	\\
13	&	Boaddoo	&	0.00798	\\
14	&	Budhiyahuttaa	&	0.00798	\\
15	&	Dhevvalaabadhoo	&	0.00798	\\
16	&	Dhevvamaagalaa	&	0.00798	\\
17	&	Dhigudhoo	&	0.00798	\\
18	&	Dhonhuseenahuttaa	&	0.00798	\\
19	&	Falhumaafushi	&	0.00798	\\
20	&	Falhuverrehaa	&	0.00798	\\
\hline
\end{tabular}
\caption{Node rank for decreasing modulus of eigenstate $\vert\psi_i\vert$ corresponding to the eigenvalue
$\lambda_2=0.97724$ (see Fig.\ref{fig6}).}
\end{table}

\begin{table}\label{table5}
\centering  
\begin{tabular}{|l|c|c|}
\hline
	&	$\lambda_{80}=-0.8165$	(``protein'')&	$\vert\psi_i\vert$	\\
\hline
\hline
1	&	Photoactivatable fluorescent protein 	&	0.22767	\\
2	&	Kaede (protein)  	&	0.13942	\\
3	&	Eos (protein)  	&	0.13942	\\
4	&	Fusion protein  	&	0.05946	\\
5	&	Green fluorescent protein 	&	0.05723	\\
6	&	Portal:Berkshire/Selected picture  	&	0.01019	\\
7	&	Persistent tunica vasculosa lentis	&	0.00552	\\
8	&	Portal:Berkshire/Selected picture/Layout  	&	0.00416	\\
9	&	Portal:Berkshire/Selected picture/1  	&	0.00416	\\
10	&	Portal:Berkshire/Nominate/Selected picture  	&	0.00416	\\
11	&	Persistent hyperplastic primary vitreous	&	0.00338	\\
12	&	Tunica vasculosa lentis 	&	0.00338	\\
13	&	Tpr-met fusion protein 	&	0.00319	\\
14	&	St Paul's Cathedral 	&	0.00256	\\
15	&	Legoland Windsor  	&	0.00255	\\
16	&	Complementary DNA  	&	0.00252	\\
17	&	Gen\'e   	&	0.00221	\\
18	&	Gene   	&	0.00215	\\
19	&	Gag-onc fusion protein 	&	0.00181	\\
20	&	Protein   	&	0.00177	\\
\hline
\end{tabular}
\caption{Node rank for decreasing modulus of eigenstate $\vert\psi_i\vert$ corresponding to the eigenvalue
$\lambda_{80}=-0.8165$ (see Fig.\ref{fig6}).}
\end{table}

\begin{table}\label{table6}
\centering  
\begin{tabular}{|l|c|c|}
\hline
	&	$\lambda_{1481}=0.1699+i0.3325$	(``Bible'')&	$\vert\psi_i\vert$	\\
\hline
\hline
1	&	Portal:Bible  	&	0.02311	\\
2	&	Portal:Bible/Featured chapter/archives 	&	0.02201	\\
3	&	Portal:Bible/Featured article 	&	0.02063	\\
4	&	Bible  	&	0.01684	\\
5	&	Portal:Bible/Featured chapter 	&	0.01644	\\
6	&	Books of Samuel	&	0.00852	\\
7	&	Books of Kings	&	0.00849	\\
8	&	Books of Chronicles	&	0.00840	\\
9	&	Book of Leviticus	&	0.00426	\\
10	&	Book of Ezra	&	0.00425	\\
11	&	Book of Ruth	&	0.00420	\\
12	&	Book of Deuteronomy	&	0.00417	\\
13	&	Book of Joshua	&	0.00400	\\
14	&	Book of Exodus	&	0.00397	\\
15	&	Book of Judges	&	0.00395	\\
16	&	Book of Genesis	&	0.00394	\\
17	&	Book of Numbers	&	0.00389	\\
18	&	Portal:Bible/Featured chapter/1 Kings	&	0.00347	\\
19	&	Portal:Bible/Featured chapter/Numbers 	&	0.00347	\\
20	&	Portal:Bible/Featured chapter/2 Samuel	&	0.00347	\\
\hline
\end{tabular}
\caption{Node rank for decreasing modulus of eigenstate $\vert\psi_i\vert$ corresponding to the eigenvalue
 $\lambda_{1481}=0.1699+i0.3325$ (see Fig.\ref{fig6}).}
\end{table}

\begin{table}\label{table7}
\centering  
\begin{tabular}{|l|c|c|}
\hline
	&	$\lambda_{1395}=-0.3149+i0.2248$	(``Austria'')&	$\vert\psi_i\vert$	\\
\hline
\hline
1	&	Lower Austria	&	0.04284	\\
2	&	Austria	&	0.03112	\\
3	&	Upper Austria	&	0.00817	\\
4	&	Styria	&	0.00781	\\
5	&	Burgenland	&	0.00307	\\
6	&	World War II	&	0.00304	\\
7	&	Krems an der Donau	&	0.00282\\
8	&	Jew	&	0.00272	\\
9	&	Slovakia	&	0.00268	\\
10	&	Bruck an der Leitha (district)	&	0.00265	\\
11	&	History of Austria	&	0.00263	\\
12	&	Wiener Neustadt	&	0.00260	\\
13	&	Mostviertel	&	0.00251	\\
14	&	States of Austria	&	0.00250	\\
15	&	Waidhofen an der Ybbs	&	0.00249	\\
16	&	MELK	&	0.00246	\\
17	&	Melk	&	0.00246	\\
18	&	Bundesland (Austria)	&	0.00239	\\
19	&	Wachau	&	0.00233	\\
20	&	Waldviertel	&	0.00226\\
\hline
\end{tabular}
\caption{Node rank for decreasing modulus of eigenstate $\vert\psi_i\vert$ corresponding to the eigenvalue
$\lambda_{1395}=-0.3149+i0.2248$ (see Fig.\ref{fig6}).}
\end{table}


\end{document}